\documentclass[11pt]{article}
\usepackage{amsmath,amsthm,amssymb,cite,enumerate}
\usepackage[a4paper,left=0.62in,right=0.62in]{geometry}

\expandafter\let\csname equation*\endcsname\relax
\expandafter\let\csname endequation*\endcsname\relax

\usepackage{amsfonts}
\usepackage{amsmath}
\usepackage{cite} 

\usepackage{color,graphicx}

\def \BE {\begin{equation}}
\def \EE {\end{equation}}
\def \BEA { \begin{eqnarray}}
\def \EEA {\end{eqnarray}}
\def \der {\nabla}

\def \pul {{\textstyle{\frac{1}{2}}}}
\def \ctvrt {\textstyle{\frac{1}{4}}}
\def \tre {\textstyle{\frac{1}{3}}}
\def \sest {\textstyle{\frac{1}{6}}}
\def \dvetre {\textstyle{\frac{2}{3}}}

\def \Mi {\stackrel{i}{M}}

\def \bl {\mbox{\boldmath{$\ell$}}}
\def \bn {\mbox{\boldmath{$n$}}}
\def \d {\mathrm{d}}

\def \hbl {\mbox{\boldmath{$\hat \ell$}}}
\def \hbn {\mbox{\boldmath{$\hat n$}}}
\def \hbm #1 {\mbox{\boldmath{$\hat m^{(#1)}$}}}
\def \bT {\mbox{\boldmath{$T$}}}

\newtheorem{thm}{Theorem}[section]

\newtheorem{prop}[thm]{Proposition}

\begin{document}

\title{Exact solutions to quadratic gravity}
\author{V Pravda$^\diamond$, A Pravdov\' a$^\diamond$, 
J Podolsk\'y$^\star$ and R \v{S}varc$^\star$\\
\vspace{0.05cm} \\
{\small $^\diamond$ Institute of Mathematics of the Czech Academy of Sciences}, \\ {\small \v Zitn\' a 25, 115 67 Prague 1, Czech Republic} \\
{\small $^\star$ Institute of Theoretical Physics, Faculty of Mathematics and Physics}, \\ {\small Charles University in Prague, V~Hole\v{s}ovi\v{c}k\'ach~2, 180~00 Praha 8, Czech Republic} \\[3mm]
 {\small E-mail: \texttt{pravda@math.cas.cz, pravdova@math.cas.cz,
}} \\ {\small \texttt{podolsky@mbox.troja.mff.cuni.cz,robert.svarc@mff.cuni.cz }}}

%

\maketitle

\abstract{
Since  all Einstein spacetimes are vacuum solutions to quadratic gravity in four dimensions, in this paper 
we study various aspects of non-Einstein vacuum solutions to this theory.   
Most such known  solutions are of traceless Ricci  and Petrov type N  with a constant Ricci scalar. Thus we assume the Ricci scalar to be constant
which leads to a substantial simplification of the field equations. 
We prove that a vacuum solution to quadratic gravity with traceless Ricci tensor of type N and aligned Weyl tensor of  any Petrov type is  necessarily a
Kundt spacetime.  This  will considerably simplify  the search for new  non-Einstein solutions. 
  Similarly, a vacuum solution to quadratic gravity with  traceless Ricci type  III and aligned Weyl tensor
of Petrov type II or more special is again necessarily a Kundt spacetime.
 
Then we study the general role of conformal transformations in constructing vacuum solutions to quadratic gravity. 
 We find that such solutions can be obtained by solving one non-linear partial differential equation for a conformal factor on any Einstein spacetime or, more generally, on any background with vanishing Bach tensor.  In particular, we show that all geometries conformal to Kundt are either Kundt or Robinson--Trautman, and we provide some explicit Kundt and Robinson--Trautman solutions to quadratic gravity by solving the above mentioned equation on certain Kundt backgrounds.}

\vspace{.2cm}

\section{Introduction and summary}

An extension of the Einstein--Hilbert action by adding  higher-order terms in curvature is a natural generalization of  Einstein's gravity theory. In the first approximation, these corrections
give quadratic terms, and in four dimensions, they admit a general form\footnote{The constant
$\gamma'$ is often denoted by $1/\kappa$.}
\BE
S = \int \d^4 x\, \sqrt{-g}\,\Big( \gamma' \left( R - 2 \Lambda \right) - \alpha' C_{abcd}\, C^{abcd} + \beta' R^2 \label{lagrQGvar} \Big)\,.
\EE
Various four and higher-dimensional theories with such quadratic terms in the action and their exact solutions were studied in the literature 
(see e.g.\cite{Gullu2011,MalPra11prd,Gurses2012,Lu2012,Baykal2014,Lu:2015cqa}), with first particular theories proposed already shortly after the introduction of Einstein's general relativity \cite{Weyl1919,Bach1921}.

The {\it vacuum}  field equations of {\it quadratic gravity} following from the action (\ref{lagrQGvar}) read
\BE
 \gamma' \left(R_{ab} - {\pul} R g_{ab} + \Lambda g_{ab} \right) -4 \alpha' B_{ab}
+2\beta' \left(R_{ab} - {\ctvrt}R g_{ab}  + g_{ab}\, \Box - \nabla_b \nabla_a \right) R
       =0 \,,
\label{EqQG4d}
\EE
where $\Box \equiv g^{ab}\, \der_a\der_b$, and $B_{ab}$ is the Bach tensor
\BE
B_{ab} \equiv \left( \nabla^c \nabla^d + {\pul} R^{cd} \right) C_{acbd} \,,
\EE
which is traceless, symmetric, and conserved (i.e., $B^{ab}_{\phantom{ab} ;b}=0$). It  can be also equivalently written as
\BE
B_{ab}={\pul}\Box R_{ab} -{\sest}\big( \nabla_a \nabla_b +{\pul} g_{ab}\Box \big)R -{\tre}RR_{ab}+R_{acbd}\,R^{cd}
+{\ctvrt}\big({\tre}R^2-R_{cd}R^{cd}\big)g_{ab} \,. \label{Bach1}
\EE

From this expression, it can be seen that the Bach tensor vanishes for all Einstein spacetimes.\footnote{Einstein spacetimes in four dimensions are defined by ${R_{ab} = \ctvrt R g_{ab}}$, where the Ricci scalar $R$ is necessarily constant.}
The last term in the field equations \eqref{EqQG4d} vanishes for Einstein spacetimes as well. This leads to the well-known observation that in four dimensions, {\it all vacuum solutions to the Einstein theory} (including possibly  a cosmological constant $\Lambda$) {\it solve also vacuum equations of the quadratic gravity} \eqref{EqQG4d}. Note that this result does not extend to dimensions ${n>4}$.
In this sense, Einstein spacetimes are trivial vacuum solutions to the four-dimensional quadratic gravity. 
The main objective of this paper is to study general properties of nontrivial solutions to quadratic gravity, i.e., {\it non-Einstein} spacetimes obeying the vacuum field equations \eqref{EqQG4d}.

 Due to the complexity of  these fourth-order nonlinear field equations, only few  non-Einstein exact solutions are known. In 1990,  non-Einstein plane wave vacuum  solutions to quadratic gravity were found \cite{Madsen90}. Recently, AdS waves admitting a cosmological constant have been constructed   using the Kerr--Schild ansatz \cite{Gullu2011}.  Note that such AdS waves, which are in fact conformal to {\it pp\,}-waves, solve quadratic gravity in any dimension. Additional non-Einstein Kundt solutions  to quadratic gravity have been found in \cite{MalPra11prd},  see section {  \ref{sec_RicciN}} for the definition of Kundt spacetimes.

In fact, all these explicit solutions to quadratic gravity are Kundt {  (i.e., spacetimes admitting a nonexpanding, shearfree, twistfree, geodetic null congruence, see sec. \ref{sec_RicciN})}. Moreover, their Ricci scalar is constant, which leads to a simplification of the field equations. In this paper,  {\it we also focus on solutions with ${R=\,}$const.} Then the trace of \eqref{EqQG4d}, which is
\BE
\gamma' \left( 4 \Lambda - R   \right)  + 6 \beta'\,  \Box R=0 \,,   \label{traceQG}
\EE
implies (for ${\gamma'\not=0}$)\footnote{Note that for certain values of the  parameters $\alpha'$, $\beta'$, $\gamma'$, the quadratic gravity reduces to more special  theories. In particular, quadratic gravity with $\beta'=0$ is {\em Einstein--Weyl gravity}. As follows from \eqref{traceQG}, for this theory the Ricci scalar is constant by default.
 Another important subcase of the quadratic gravity is {\em conformal gravity} given by ${\beta'=0=\gamma'}$, for which the field equations \eqref{EqQG4d} reduce to ${B_{ab}=0}$. }
 that  
\BE
{R=4\Lambda}\, , \label{Ricciconst}
\EE
and the field equations \eqref{EqQG4d}
reduce considerably  to
 \BE
(\gamma'+8\beta'\Lambda)(R_{ab}-\Lambda g_{ab})=4\alpha' B_{ab}\,.  \label{eqQGRconst}
\EE

\subsection{Kundt solutions to quadratic gravity}
Since for all the above mentioned solutions to quadratic gravity their Weyl and traceless Ricci tensors are of type N in the algebraic classification 
\cite{Coleyetal04,OrtPraPra12rev},
we begin with vacuum solutions to quadratic gravity  with traceless Ricci tensor of type N. In section \ref{sec_Kundt}, we will prove
\begin{prop}
\label{prop_RicciN}
A vacuum solution to quadratic gravity \eqref{EqQG4d} with the Ricci tensor of the form  
$$
R_{ab}=\Lambda g_{ab}+\omega'\ell_a\ell_b\,,  \ \ \ \omega' \not=0\,, \ \ \ {  \ell^a\ell_a=0}\,,\ \ \label{RicciN}
$$
and aligned Weyl tensor of any Petrov type is necessarily Kundt.
\end{prop} 
For traceless  Ricci type N and aligned Weyl  type N, this result has been already obtained in \cite{MalPra11prd}.
Note that in contrast, for $\omega'=0$ (Einstein spacetimes), expansion and twist (and for Petrov type I also
shear) can be nonvanishing.

Then, we  will proceed with a generalization of this result to the case of traceless Ricci type III:
\begin{prop}
\label{prop_RIII_Kundt}
A vacuum solution to quadratic gravity \eqref{EqQG4d} with the Ricci tensor of the form  
$$
R_{ab}=\Lambda g_{ab}+\psi'_i (\ell_a  m^{(i)}_b+  m^{(i)}_a\ell_b)+\omega'\ell_a\ell_b\,, \ \   \psi'_i \psi'_i  \not=0\,,\
$$
and aligned Weyl tensor of Petrov type II, or more special, is necessarily Kundt.
\end{prop}
{  Note that the Ricci tensor is expressed using a null frame introduced in sec. \ref{sec_Kundt}.}

Since the Kundt spacetimes have been extensively studied (see \cite{Stephanibook,GriffithsPodolsky:2009}), propositions \ref{prop_RicciN} and \ref{prop_RIII_Kundt} will allow for a systematic search of vacuum solutions of quadratic gravity \eqref{EqQG4d} of the above Ricci types. Particular examples of such Kundt solutions \cite{Madsen90,Gullu2011,MalPra11prd}  were mentioned  above.

\subsection{Conformally Kundt solutions to quadratic gravity}

Non-Kundt (and non-Einstein) solutions to quadratic gravity  also exist. 
Remarkably, a non-Schwarzschild static spherically symmetric black hole solution in Einstein--Weyl gravity (with $\Lambda = 0$) has been found very recently   in \cite{Lu:2015cqa}, where its two metric functions are given in terms of two ODEs. 
We will point out that this solution belongs to the Robinson--Trautman {  (RT)} class and in fact 
due to (the {  part} 
of) proposition \ref{prop_conformal} of section  \ref{sec_KundtRT} it is necessarily conformal to Kundt:
\begin{prop}
All Robinson--Trautman spacetimes are conformal to Kundt.
\end{prop}

Under the conformal transformation
\BE
{\tilde {g}}_{ab} = \Omega^2  g_{ab}\,, \label{conftrans}
\EE
the Bach tensor transforms as
\BE
{\tilde B}_{ab} = \Omega^{-2} B_{ab}\,. \label{confBach}
\EE
Thus obviously, the Bach tensor vanishes not only for all Einstein spacetimes but also for all spacetimes {\it conformal} to Einstein spacetimes. However, vanishing of the Bach tensor is  not a sufficient condition for a spacetime  to be conformally related to an Einstein spacetime \cite{Kozameh85}. Indeed, explicit examples of spacetimes with vanishing Bach tensor which are not conformal to Einstein spacetimes are known \cite{Bergqvist2007,Liu2013}.

One can  employ  \eqref{confBach} to construct new exact solutions to quadratic gravity with arbitrary nonzero parameters $\alpha'$, $\beta'$, $\gamma'$ but a special value of the cosmological constant
\BE
\Lambda = - \frac{\gamma'}{8 \beta'}\, 
\label{lambdaspec}
\EE
with \eqref{Ricciconst}.
The case ${\Lambda \not= - \frac{\gamma'}{8 \beta'}}$ will be discussed elsewhere.
Under the assumption \eqref{lambdaspec}, the equations of quadratic gravity \eqref{eqQGRconst} reduce to
\BE
B_{ab} = 0\,.  \label{eqQGspecLambda}
\EE

Specifically, we will use ``seed'' geometries $g_{ab}$ with  vanishing Bach tensor {\it to  generate solutions ${\tilde g}_{ab}$ to quadratic gravity using  the conformal transformation \eqref{conftrans}}, implying ${{\tilde B}_{ab}=0}$ due to \eqref{confBach}.\footnote{Note that the seed metrics themselves {\it need not} to solve the field equations \eqref{EqQG4d} of quadratic gravity.} It remains to satisfy 
\BE
\tilde{R}  = 4 \Lambda\,, \label{confRicciconst} 
\EE
where $\tilde{R}$ is the Ricci scalar of the conformally transformed metric ${\tilde {g}}_{ab} $.
It is well known that the Ricci scalars of conformally related metrics obey 
\BE
{\tilde R = R\, \Omega^{-2} - 6\, \Omega^{-3}\, \Box \Omega}\,.
\EE
Thus we can satisfy \eqref{confRicciconst} by choosing the conformal factor $\Omega$ that solves the equation
\BE
6\, \Box \Omega - R\, \Omega +  4\Lambda\,\Omega^3 = 0\,,  \label{eqOmR4d}
\EE
where $\Lambda$ is constrained by \eqref{lambdaspec}.
Thus, in this approach,  the problem of constructing new solutions to quadratic gravity reduces
to solving one nonlinear PDE \eqref{eqOmR4d} with a cubic nonlinearity  for one unknown function $\Omega$ on a curved background spacetime with  vanishing Bach tensor. 

In section~\ref{sec_exact}, we will use this generating technique to derive several explicit new solutions to quadratic gravity.

To conclude, let us note that there are solutions to quadratic gravity that are neither Kundt nor Robinson--Trautman.
One such Petrov type N twisting solution will be briefly discussed
in section~\ref{sec_exact}. For this solution, the Ricci tensor is more general than the form given in proposition \ref{prop_RIII_Kundt}.

\section{Kundt spacetimes in quadratic gravity}
\label{sec_Kundt}

In this section, we study spacetimes with certain algebraically special forms of the Ricci tensor and aligned Weyl tensor. 
We show that the vacuum field equations of quadratic gravity imply that these spacetimes are necessarily Kundt.

In the case of constant Ricci scalar, the Bach tensor \eqref{Bach1} can be expressed as
\BE
B_{ab}={\pul}\Box R_{ab}  -{\tre}RR_{ab}+R_{acbd}\,R^{cd}
+{\ctvrt}\big({\tre}R^2-R_{cd}R^{cd}\big)g_{ab} = B_{ab}^R+B_{ab}^C\,, \label{Bach3}
\EE
where
\BEA
B_{ab}^R&=&{\pul}\Box R_{ab}  +{\tre}RR_{ab}-R_{ac}\,R^{c}_{\ b}
+{\ctvrt}\big(R_{cd}R^{cd}-{\tre}R^2\big)g_{ab}\,, \label{BachR3}\\
B_{ab}^C&=&C_{acbd} R^{cd}
\EEA
are  parts of the Bach tensor depending only on the Ricci tensor and also on the Weyl tensor, respectively.
Let us employ a real null frame with two null vectors $\bl$ and $\bn$ and two spacelike vectors $\mbox{\boldmath{$m^{(i)}$}} $ ({$i,j =2,3$}) obeying
\BE
\ell^a \ell_a= n^a n_a = 0, \qquad   \ell^a n_a = 1, \qquad \ m^{(i)a}m^{(j)}_a=\delta_{ij}\,.  \label{ortbasis}
\EE
For the algebraic classification of tensors, the crucial concept is a {\it boost weight} {$($b.w.$)$}.
A quantity  $q$  has  the boost weight   ${\rm b}$  if it transforms as
\BE
\hat q = \lambda^{\rm b} q
\EE
under a boost
\BE
\hbl = \lambda \bl, \qquad    \hbn = \lambda^{-1} \bn, \qquad   \hbm{i} = \mbox{\boldmath{$m^{(i)}$}}  \label{boost}\,.
\EE

Various frame components of a tensor will have in general different integer boost weights and we define boost order of a tensor $\bT$ with respect to a given frame as the maximum b.w. of its frame components. It can be shown that the boost order of $\bT$ in fact depends only on the frame vector $\bl$ and thus  we will denote it as ${\rm b}_{\bl }(\bT)$ (see,
e.g., \cite{OrtPraPra12rev}). Obviously ${\rm b}_{\bl }(\bT_1 \otimes \bT_2) = {\rm b}_{\bl }(\bT_1) + {\rm b}_{\bl }(\bT_2)$. Note also that boost order of  a  tensor does not increase under a contraction of indices.

Let us study spacetimes with the Ricci tensor of the form
\BE
R_{ab}=\Lambda g_{ab}+\psi'_i (\ell_a  m^{(i)}_b+  m^{(i)}_a\ell_b)+\omega'\ell_a\ell_b \,,\label{Ricci}
\EE
which clearly obeys \eqref{Ricciconst}.
Boost order of the traceless Ricci tensor is thus $-1$ (for $\psi'_i \psi'_i \not= 0$) or $-2$ (for {$\psi'_i\psi'_i=0$,} $\omega'\not=0$) and thus the Ricci tensor is 
of type III or N, respectively (see \cite{OrtPraPra12rev}). 
From \eqref{BachR3}, we obtain
\BE
B_{ab}^R={\pul}\Box R_{ab}-\dvetre \Lambda \psi'_i (\ell_a  m^{(i)}_b+  m^{(i)}_a\ell_b)
-\left(\dvetre \Lambda\omega' + \psi'_i\psi'_i\right) \ell_a\ell_b\,.\label{BR}
\EE

\subsection{Traceless Ricci type N}
\label{sec_RicciN}

First, let us focus on the traceless Ricci type N for which
\begin{equation}
R_{ab}=\Lambda g_{ab}+\omega' \ell_a \ell_b,\ \  \omega'\not=0 \,. \label{RiccitypeN}
\end{equation}
From the contracted Bianchi equations $\nabla^b R_{ab}= \frac{1}{2} \nabla_a R=0$ and \eqref{RiccitypeN},  it follows that $\bl$ is geodetic 
and 
without loss of generality, one can choose $\bl$ to be affinely parametrized and a frame to be parallelly transported along $\bl$.
Then, the covariant derivatives of the frame vectors in terms of spin coefficients read   \cite{OrtPraPra12rev}
\BEA
\ell_{a ; b} &=& L_{11} \ell_a \ell_b  + L_{1i} \ell_a m^{(i)}_{\, b}  +
\tau_i 
m^{(i)}_a \ell_b    + {\rho}_{ij} m^{(i)}_{\, a} m^{(j)}_{\, b} \, , \label{dl} \\
n_{a ; b  } &=&\! -\! L_{11} n_a \ell_b -\! L_{1i} n_a m^{(i)}_{\, b}  +
\kappa'_i 
 m^{(i)}_{\, a} \ell_b   + \rho'_{ij} 
m^{(i)}_{\, a} m^{(j)}_{\, b}  \,, \label{dn} \\
m^{(i)}_{a ; b } &=&\! -\! \kappa'_i 
\ell_a \ell_b  -\tau_i 
n_a \ell_b  
-\rho'_{ij} 
\ell_a m^{(j)}_{\, b}   
 \! +\! {\Mi}_{j1} m^{(j)}_{\, a} \ell_b  -\rho_{ij} n_a m^{(j)}_{\, b} 
+ {\Mi}_{kl} m^{(k)}_{\, a} m^{(l)}_{\, b} \,. \label{dm} 
\EEA
{  Here, the optical matrix
\BE
\rho_{ij}\equiv \ell_{a ; b}m_{(i)}^a m_{(j)}^b
\EE
can be decomposed into its trace $\theta$  {(\em expansion)}, 
trace-free symmetric part $\sigma_{ij}$ and antisymmetric part $A_{ij}$, namely
\BEA
 \rho_{ij}=\sigma_{ij}+\theta\delta_{ij}+A_{ij} , \label{L_decomp} \qquad 
 \sigma_{ij}\equiv \rho_{(ij)}-\textstyle{\frac{1}{2}} \rho_{kk}\delta_{ij} , 
\qquad \theta\equiv\textstyle{\frac{1}{2}}{\rho_{kk}} , \qquad A_{ij}\equiv \rho_{[ij]} . \label{opt_matrices}
\EEA
Optical scalars {\em shear} and {\em twist} of $\bl$ are traces  $\sigma^2\equiv\sigma^2_{ii}=\sigma_{ij}\sigma_{ji}$ and $\omega^2\equiv-A^2_{ii}=-A_{ij}A_{ji}$,
respectively.
Kundt spacetimes are defined as spacetimes with vanishing $\rho_{ij}$.
}

Using \eqref{RiccitypeN} and \eqref{dl}, we express $\der_c R_{ab}$
\BE
\der_c R_{ab}= D \omega' \ell_a\ell_b n_c +\omega' \rho_{ij}(m^{(i)}_a \ell_b+\ell_a m^{(i)}_b)m^{(j)}_c
+\mbox{terms of b.w. $\leq-2$}\,,
\label{dR}
\EE
where $D\equiv \ell^a\der_a$.
Employing \eqref{dl}--\eqref{dm},  
a further differentiation of \eqref{dR} leads to
\BE
\Box R_{ab}=[-\omega' \rho_{ij}\rho_{ij}(\ell_a n_b+n_a\ell_b)+2\omega'\rho_{ik}\rho_{jk} m^{(i)}_a m^{(j)}_b]+\mbox{terms of b.w. $\leq-1$}\,,
\label{boxRN}
\EE
i.e., from \eqref{BR} and \eqref{boxRN}
\BE
B^R_{ab}=\pul [-\omega' \rho_{ij}\rho_{ij}(\ell_a n_b+n_a\ell_b)+2\omega'\rho_{ik}\rho_{jk} m^{(i)}_a m^{(j)}_b]+\mbox{terms of b.w. $\leq-1$}\,.\label{BRN}
\EE
For the Ricci tensor of the form \eqref{RiccitypeN}, the left-hand side of the field equations of quadratic  gravity \eqref{eqQGRconst} contains b.w. $-2$ terms
 only,
while in general, the right-hand side  contains terms of b.w. $0$ due to the presence of the term $\Box R_{ab}$. Recall that we assume that the Weyl tensor is aligned with the Ricci tensor (i.e.  ${\rm b}_{\bl} (C_{abcd}) \leq 1$)\footnote{Note that for the Weyl types III and N this is not an assumption since from the Bianchi equations it follows 
that Weyl type III/N traceless Ricci type N spacetimes are aligned (see \cite{wils}).} and thus 
${\rm b}_{\bl }(B^C_{ab}) ={\rm b}_{\bl }(C_{abcd}) +{\rm b}_{\bl }(R_{ab}-\Lambda g_{ab}) \leq  -1$.
Consequently, the leading term in \eqref{BRN} has to vanish, i.e., 
\BE
\rho_{ij} \rho_{ij} = 0\,.
\EE
Thus the optical matrix  vanishes, $\rho_{ij}=0$, obviously implying also $\rho_{ik} \rho_{jk} = 0$ for all $i,j$,
which  concludes the proof of proposition \ref{prop_RicciN}.

\subsection{Traceless Ricci type III}

Let us proceed with a more general form of the Ricci tensor \eqref{Ricci} with $\psi'_i\psi'_i\not=0$. First, we prove that  $\bl$ is geodetic using the standard four-dimensional Newman--Penrose (NP) formalism.
For the Ricci tensor of the form \eqref{Ricci}, the relevant NP components are $\Phi_{22}$ and $\Phi_{12}={\bar \Phi_{21}}\not=0$.
For the Petrov types III/N/O, the  Bianchi equation (7.32b) of  \cite{Stephanibook} reduces to
\BE
\kappa\, \Phi_{12} = 0\,, 
\EE
which implies that $\bl$ is geodetic. Similarly, for the Petrov type  II, 
equation (7.32a) of  \cite{Stephanibook} gives 
$\kappa \Psi_2 = 0$ and thus $\bl$ is also geodetic.

The first derivative of the Ricci tensor \eqref{Ricci} reads
\BEA
\der_c R_{ab}&=&  m^{(i)}_a m^{(j)}_b m^{(k)}_c (\psi'_j \rho_{ik}+\psi'_i\rho_{jk}) 
+(m^{(i)}_a \ell_b+\ell_a m^{(i)}_b)n_c D\psi'_i\nonumber\\
&& +(n_a \ell_b+\ell_a n_b)m^{(i)}_c (-\psi'_s\rho_{si}) + {\rm b.w.} \leq -1 \ {\rm terms}\,.
\EEA
Further differentiation gives
\BE
\Box R_{ab}=-(m^{(i)}_a n_b+n_a m^{(i)}_b)(2 \psi'_s\rho_{sk}\rho_{ik}+\psi'_i\rho_{sk}\rho_{sk})+ \ \mbox{terms of b.w. $\leq 0$}\,,\label{BoxRIII}
\EE
i.e. from \eqref{BR} and  \eqref{BoxRIII}
\BE
B^R_{ab}=-\pul (m^{(i)}_a n_b+n_a m^{(i)}_b)(2 \psi'_s\rho_{sk}\rho_{ik}+\psi'_i\rho_{sk}\rho_{sk})+ \ \mbox{terms of b.w. $\leq 0$}\,.\label{BRIII}
\EE
If the Weyl tensor of any Petrov type  and the Ricci tensor \eqref{Ricci} are aligned then ${\rm b}_{\bl }(B^C_{ab}) \leq  0$.
Thus the b.w. $+1$ terms in \eqref{BRIII} are the only  b.w. $>0$ terms in \eqref{eqQGRconst} and therefore they  have to vanish.
By multiplying \eqref{BRIII} by $\psi'_i$, we get
\BE
(\psi'_i\psi'_i) (\rho_{sk}\rho_{sk}) +2 (\psi'_s\rho_{sk})(\psi'_i\rho_{ik})=0\label{Bbw1}\,.
\EE
For $\psi'_i \psi'_i \not=0$, the expression \eqref{Bbw1} clearly vanishes iff $\rho_{ij}=0$ which concludes the proof of 
proposition \ref{prop_RIII_Kundt}.

Note that proposition \ref{prop_RIII_Kundt} is valid also for aligned Petrov type I spacetimes with the Ricci tensor of the form \eqref{Ricci} and a geodetic $\bl$, however, in this case we did not prove geodecity of $\bl$.

\section{Conformal relations of Kundt and Robinson--Trautman spacetimes}
\label{sec_KundtRT}

{  To our knowledge, 
all exact solutions to quadratic gravity discussed in the literature } so far are either Kundt, or conformal to it. It is thus important to identify the class of all spacetimes that are conformal to Kundt geometries, which admit a null geodetic congruence~$\bl$  with vanishing  shear, twist and expansion.\footnote{Since the results of this section are dimension-independent, here
we work in a general dimension $n$.} It has a canonical metric form \cite{Kundt:1961,Coleyetal03,ColHerPel06,PodolskySvarc:2015a,PodZof09}
\BE
 \d s^2_{\rm Kundt} =2H(u,r,x)\,\d u^2-2\,\d u\d r+2W_i(u,r,x)\,\d u\d x^i+ g_{ij}(u,x)\, \d x^i\d x^j \,, \label{Kundt_gen}
\EE
where ${\bl=\partial_r}$ (with a dual ${-\d u}$). A generic Kundt metric is of the Riemann type I or more special \cite{OrtPraPra07,
PodZof09}, and of the Weyl subtype I(b) in $n>4$ \cite{PodolskySvarc:2013a}, with $\bl$ being both {  principal null direction (PND)} and an aligned null direction of the Ricci tensor (since ${R_{rr}}$ vanishes identically).

Now, a conformally transformed metric
\BE
\d {\tilde s}^2 = \Omega^2 (u,r,x)\, \d s^2_{\rm Kundt} \label{confKundt}
\EE
is obviously of the {\it same Weyl type}, while the {\it Ricci type is in general distinct} from  the Ricci type of \eqref{Kundt_gen}.
The vector $\bl$ is also a null geodetic {  PND} of the  new metric \eqref{confKundt} with  vanishing  shear and twist, while its nontrivial {\it expansion} reads
\BE
\tilde \theta = \frac{1}{n-2}\, ({\tilde g}^{ab}\ell_{b})_{;a} = \frac{\Omega_{,r}}{\Omega^3}\,. \label{eqtheta}
\EE
Under the conformal transformation \eqref{confKundt}, the Ricci tensor and scalar transform as \cite{Wald}
\BEA
{\tilde R}_{ab}&=&R_{ab}-\Omega^{-1}\big[(n-2)\delta^c_a\delta^d_b+g_{ab}g^{cd}\big]
    \,\nabla_c\nabla_d\Omega
\nonumber\\
&& +\Omega^{-2}\big[2(n-2)\delta^c_a\delta^d_b-(n-3)g_{ab}g^{cd}\big](\nabla_c\Omega)(\nabla_d\Omega)\,,\\
\ \tilde R &=& R\, \Omega^{-2} - 2(n-1)\Omega^{-3} \Box \Omega-(n-1)(n-4) \Omega^{-4} (\nabla_a \Omega) (\nabla^a \Omega) \,.  \label{eqOm}
\EEA
Thus, in contrast with the Kundt metric \eqref{Kundt_gen},  for the new metric \eqref{confKundt}, {\it the highest boost weight component of the Ricci tensor is in general nonvanishing\,}:\footnote{We set ${\tilde \ell}_a=\ell_a$. Note that this choice preserves geodeticity and the affine parametrization.
}   
\BE
{\tilde R}_{ab} \,{\tilde \ell}^a {\tilde \ell}^b =
{\tilde g}^{ac}{\tilde g}^{bd}{\tilde R}_{ab}\, \ell_c \ell_d = \Omega^{-4} {\tilde R}_{rr}
 = - (n-2)\,\Omega^{-6} (\Omega \Omega_{,rr}-2 \Omega_{,r}^2) \,.
\EE

Since ${\bl=\partial_r}$ is geodetic, shearfree and twistfree null direction in the conformally related metric \eqref{confKundt}, this new metric  is a {\it Robinson--Trautman\,} metric (as long as ${\Omega_{,r} \not= 0}$) or {\it Kundt\,} for ${\,\Omega_{,r} = 0\Leftrightarrow\tilde \theta = 0}$, see \eqref{eqtheta}. This can  be explicitly seen by transforming \eqref{confKundt} into the canonical Robinson--Trautman form \cite{RobTra60,PodOrt06,PodolskySvarc:2015a}
\BE
 \d {\tilde s}^2_{\rm RT}=2{\tilde H}(u,{\tilde r},x)\,\d u^2-2\,\d u \d {\tilde r}+2{\tilde W}_i(u,{\tilde r},x)\,\d u\d x^i+{\cal R}^2 (u,{\tilde r},x) g_{ij}(u,x)\,\d x^i\d x^j \,,
 \label{RT_metric}
\EE
where
\BEA
{\tilde r}&=&\rho(u,r,x),\,\quad \mbox{such that}\quad   
\rho_{,r}=\Omega^2(u,r,x)\,,\  
\nonumber\\ 
\d {\tilde r}&=&\Omega^2\, \d r+ \rho_{,u}\,\d u+ \rho_{,i}\, \d x^i\,, \nonumber\\
{\tilde H}&=&\Omega^2   H+\rho_{,u} \,,\quad \nonumber\\
{\tilde W}_i &=& \Omega^2   W_i +\rho_{,i} \,,\quad\nonumber\\
{\cal R} &=&\Omega \,. \label{T1}
\EEA

In fact, by comparing the expansion, ${\theta=\textstyle{\frac{1}{n-2}}\,\ell^a_{\;;a}\, }$, shear
${\sigma^2=\ell_{(a;b)}\ell^{(a;b)}-\textstyle{\frac{1}{n-2}}\left(\ell^a_{\;;a}\right)^2\,}$, and twist ${\omega^2=\ell_{[a;b]}\ell^{a;b}\,}$ 
of the geodetic affinely parametrized null vector $\ell_a$ expressed in the original and conformally transformed metrics,
we arrive at (see also \cite{RobTra83} for conformal properties of flows in arbitrary  dimension)
\BE
\tilde{\theta} = \frac{\theta}{\Omega^2} + \frac{\bl(\Omega)}{\Omega^3}\,, \qquad {\tilde \sigma}^2 = \frac{\sigma^2}{\Omega^4}\,, \qquad {\tilde \omega}^2 = \frac{\omega^2}{\Omega^4}\,,
\EE
{  where $\bl(\Omega)\equiv\Omega_{,r}$}. 
Together with the above results, this leads to
\begin{prop}
\label{prop_conformal}
{Spacetimes conformal to shearfree or twistfree spacetimes are shearfree or twistfree, respectively. In particular:}
\begin{enumerate}
\item
All spacetimes conformal to Robinson--Trautman are Robinson--Trautman or Kundt.
\item
All Robinson--Trautman spacetimes are conformal to Kundt.
\item
All spacetimes conformal to Kundt are  Robinson--Trautman (when ${\bl(\Omega)\equiv\Omega_{,r}  \not=0}$) {  or} Kundt (when ${\bl(\Omega) \equiv \Omega_{,r} =0}$).\footnote{{Since Kundt and Robinson--Trautman spacetimes are defined by the existence of a geodetic shear-free and twist-free null congruence with 
$\theta=0$ and $\theta\not=0$, respectively, interestingly there are exceptional spacetimes that belong to {\it both} of these classes admitting two distinct congruences with these properties, see, e.g., the metric \eqref{nonKundtmetric}, \eqref{conRTsubcase}.}} 
\end{enumerate}
\end{prop}

Note that in the case of four dimensions, an extension of the Goldberg--Sachs theorem to conformally Einstein spacetimes immediately follows \cite{Stephanibook} and thus algebraically special solutions to quadratic  gravity obtained by  a conformal transformation of Einstein spacetimes are shearfree.

It has been shown \cite{Pravdaetal04,PodOrt06} that in contrast to the four-dimensional case, for ${n>4}$ {\it Einstein} Robinson--Trautman spacetimes of types III and N do not exist. From the above results, it follows that {\it non-Einstein  Robinson--Trautman geometries of types N and III can be clearly constructed by a conformal transformation from their Kundt counterparts in any dimension}. Furthermore, starting with universal Kundt metrics of types N and III \cite{HerPraPra14}, one obtains type N and III Robinson--Trautman vacuum solutions to ${n>4}$ conformal gravities {  (theories of gravity invariant under conformal transformations)}. The strong constraints on the optical matrix of higher-dimensional type N and III spacetimes, implying the nonexistence of Einstein Robinson--Trautman solutions within these classes, is thus connected to the Einstein field equations  rather   than to the geometric properties of Robinson--Trautman spacetimes in higher dimensions.

\subsection{Static spherically symmetric spacetimes}

As an important illustration, let us investigate static spherically symmetric spacetimes
\BE
\d s^2= -h({\bar r})\, \d t^2+\frac{\d {\bar r}^2}{f({\bar r})}
+{\bar r}^2\,\d \omega^2_{n-2}\,, \qquad
\d \omega^2_{n-2}=(1+{\ctvrt}\delta_{kl}x^kx^l)^{-2}\,\delta_{ij}\,\d x^i\d x^j\,.
\label{statspher}
\EE
These spacetimes are of the Weyl type D in any dimension \cite{PraPraOrt07} and include many black hole solutions of various theories. They belong to the Robinson--Trautman class. Indeed, by performing  a coordinate transformation
\BE
\d t=\d u+\frac{\d {\bar r}}{\sqrt{hf}}\,, \qquad
\d {\tilde r} = \sqrt{{h}/{f}}\,\d {\bar r}\label{rbarrtilde}
\EE
we arrive at the {\it canonical Robinson--Trautman form} \eqref{RT_metric} with ${{\tilde W}_i=0}$
\BE
\d {\tilde s}^2_{\rm RT}=-h\big({\bar r}({\tilde r})\big)\,\d u^2-2\,\d u \d {\tilde r}+{\bar r}^2({\tilde r})\,\d \omega^2_{n-2}\,. \label{SSmetRT}
\EE
A further coordinate transformation \eqref{T1} for ${{\tilde r}=\rho(r)}$ such that
${\,\d {\tilde r}=\Omega^2(r)\, \d r\,}$
brings the metric \eqref{SSmetRT} to the {\it form manifestly conformal to Kundt}
\BE
\d {\tilde s}^2_{\rm RT}=\Omega^2(r)\, \d s^2_{\rm Kundt}
=\Omega^2(r) \,\big({\cal H}(r)\,\d u^2 - 2\, \d u \d r+\d \omega^2_{n-2}\big), \label{SSCFKundt}
\EE
cf. \eqref{Kundt_gen} with ${{\cal H}=2H}$ and ${W_i=0}$,  where
\BEA
\Omega^2 \,{\cal H}&=&-h\,, \nonumber\\ 
\Omega(r)&=&{\bar r}\big({\tilde r}(r)\big),\,  
\nonumber\\
\sqrt{\frac{f}{h}}
&=&\frac{\d {\bar r}}{\d {\tilde r}}
=\frac{\d \Omega}{\d {\tilde r}}
=\Omega_{,r}\,\frac{\d r}{\d {\tilde r}}
=\frac{\Omega_{,r}}{\Omega^2}\,.\label{eq-hOm}
\EEA
Note that the ``seed'' Kundt metric appearing in \eqref{SSCFKundt} is a {\it direct-product geometry}, containing as a special cases, e.g., Bertotti--Robinson or Nariai space \cite{GriffithsPodolsky:2009,KadlecovaZelnikovKrtousPodolsky:2009}.

For the Schwarzschild--Tangherlini solution, 
\BE
{f({\bar r})=h({\bar r})=1-\mu\, {\bar r}^{\,3-n}}
\EE
 in \eqref{statspher}, and thus 
\BEA
\Omega&=&{\bar r}={\tilde r}={- 1/r}, \nonumber\\
{\cal H}&=&(-1+\mu\, {\bar r}^{\,3-n})/{\bar r}^2=-r^2+{  (-1)^{n-1}\mu} 
\,r^{n-1}.
\EEA

\section{Exact solutions to quadratic gravity}
\label{sec_exact}

Now let us discuss solutions to quadratic gravity \eqref{eqQGspecLambda} obtained via solving equation \eqref{eqOmR4d} on an appropriate seed spacetime.  In principle, one can solve this nonlinear equation numerically on any background Einstein spacetime or, more generally, on any spacetime with  vanishing Bach tensor. Here we will focus on cases where solutions can be obtained explicitly.

First, let us note that \cite{Liu2013} gives a list of several metrics with vanishing Bach tensor that are not conformally Einstein. Many of them have constant Ricci scalar. Thus, one can obtain exact solutions of quadratic gravity \eqref{eqQGspecLambda} from these seeds by
appropriate constant rescaling of these metrics to set  ${\tilde R = 4 \Lambda = - \frac{\gamma'}{2 \beta'}}$. Interestingly, apart from Kundt metrics (solutions 2--5 of \cite{Liu2013}), this rescaling leads also to solutions of quadratic gravity outside the Kundt and Robinson--Trautman classes. For example
for a type N twisting, shearfree, expansion-free metric 6  of \cite{Liu2013} \footnote{{ {  The multiple PND,} 
$\bl={\partial_v}$, of the metric \eqref{twisting} is twisting and thus it is not Kundt, nor RT. Since for Petrov type N the PND is unique this metric also does not admit a Kundt or RT congruence distinct from $\bl$ (it follows from purely geometric considerations
that  Kundt or RT congruences always coincide with PNDs \cite{OrtPraPra12rev}, \cite{PodZof09}, \cite{PodSvRT}).}}
\BE
\d s^2=\frac{\d r^2}{r^2}+10\d u\left( \frac{\d v}{r^2}-\frac{2\d x}{r}\right)
+2r\d x\d v+\frac{10\d u^2}{r^4}+r^2\d x^2, \label{twisting}
\EE
the Ricci tensor is  constant, $R=-3$, and thus an appropriate constant rescaling leads to a type N twisting solution of quadratic gravity with $\gamma'/\beta'>0$.

In the following, we focus on solving  equation \eqref{eqOmR4d} on the Kundt backgrounds \eqref{Kundt_gen}. {This approach leads to Kundt and Robinson-Trautman solutions of quadratic gravity, cf. proposition \ref{prop_conformal}}. The d'Alembert operator applied to a function $\Omega(u,r,x^{i})$ then reads explicitly
\BEA
\Box \Omega \equiv g^{ab}\,\Omega_{;ab}
&=&
(-2H+W^i W_i)\,\Omega_{,{rr}} -2\Omega_{,{ru}} +2W^i\, \Omega_{,{ri}}+g^{ij}\,\Omega_{||ij} \nonumber
\\ & &
+\big[\!-2 H_{,r} +2W^iW_{i,r}  +g^{ij} (W_{(i||j)}-{\pul} g_{ij,u})  \big]\,\Omega_{,r}
+g^{ij}\,W_{i,r}\,\Omega,_j\,,
\label{box_genKundt}
\EEA
where $||$ denotes the covariant derivative associated with the spatial metric $g_{ij}$.
In case of Kundt metrics without off-diagonal terms ($g_{ui}=W_i=0$),  this
simplifies considerably to
\BE
\Box \Omega=-2H\,\Omega_{,rr} -2\,\Omega_{,ru}-2H_{,r}\,\Omega_{,r}
+g^{ij} ( \Omega_{||ij} -{\pul} g_{ij,u}\,\Omega_{,r})\,. \label{box_diagKundt}
\EE

\subsection{Solutions generated by a {\it pp}-wave seed}
\label{sec_ppseed}

On a {\it pp\,}-wave background,
\BE
\d s^2_{\rm seed}= 2H(u,x,y)\, \d u^2 -2\,\d u \d r+ \d x^2+\d y^2\,, \label{pp}
\EE
equation \eqref{eqOmR4d} with \eqref{box_diagKundt}, {  assuming $\Omega$ to be function of $z$, where} 
\BE
z \equiv p_a \, x^a=p_r\, r+p_u\, u+p_x\, x+p_y\, y\,, \ {  p_a=\, \mbox{const.}}
\EE
and {  using} the fact that ${R=0}$, reduces to
\BE
\Omega'' +\frac{{\tilde R}}{6p_ap^a} \,\Omega^3=0\,,\label{eqOmega}
\EE
where prime denotes the derivative with respect to $z$ {  and the contravariant vector $p^a$ is obtained using the metric \eqref{pp}}.
If either ${p_r=0}$ or ${H=\,}$const., this is a constant-coefficients ODE for ${\Omega(z)}$, with the first integral
\BEA
{\Omega'}^2&=&-L \,\Omega^4+K\,, \nonumber\\
K&=&\mbox{const.},\qquad L={\frac{{\tilde R}}{12p_ap^a}}
=\frac{{\Lambda}}{3p_ap^a}\,.\label{odeOm}
\EEA
Equation \eqref{odeOm} can be integrated:
\begin{itemize}
\item
for the case with a vanishing integration constant ${K=0}$ {  and $L<0\,$,} 
\BE
\Omega(z)=\pm\sqrt{-\frac{1}{L} }\, \frac{1}{z-c}\,,\qquad c=\mbox{const.}\,, \label{solOmegaspec}
\EE
\item
while for a nonvanishing $K$ it can  be solved in terms of elliptic Jacobi function\footnote{Note that sometimes this elliptic function is denoted as ${\,{\rm sn}(\quad, {\rm i})\,}$.} 
\BE
\Omega(z)= K^{\frac{1}{4}}L^{-\frac{1}{4}}\, {\rm sn} \big( K^{\frac{1}{4}}L^{\frac{1}{4}} (z-c), -1 \big)\, , \ \ \ c=\mbox{const.}\,, \ \ 
{  K>0\,,\ \ L\not=0\,. }\label{sn_sol}
\EE
{In the $L\ \rightarrow\ 0$ limit, this reduces to
\BE
\Omega(z)= \sqrt{K} (z-c)\, , \ \ \ c=\mbox{const.}\,, \ \ K>0\,, \label{Lnul}
\EE
which solves \eqref{eqOmega} with ${\tilde R}=0$.
However,  note that \eqref{Lnul} corresponds to a solution  
of a special subcase of quadratic gravity with $\gamma'=0$ only.}

\end{itemize}

In general, the Weyl, Ricci, and Bach tensors  of \eqref{pp} possess only boost weight ${-2}$ components. In particular
\BEA
R_{uu} &=& -\Delta H=- (H,_{xx}+H,_{yy})\,, \\
B_{uu} &=& -{\pul}\Delta\Delta H=-{\pul} (H,_{xxxx}+2H,_{xxyy}+H,_{yyyy})\,,
\EEA
so that $R_{uu}$ and $B_{uu}$ vanish if, and only if,
\BEA
R_{uu}=0\ \ \Leftrightarrow\ \ H &=& F(\zeta)+\bar F(\bar\zeta)\, ,\\
B_{uu}=0\ \ \Leftrightarrow\ \ H &=& F(\zeta)+\bar F(\bar\zeta)+ \bar\zeta\, G(\zeta)+\zeta\, \bar G(\bar\zeta)\, , \label{Bach0H}
\EEA
respectively, where $F$ and $G$ are arbitrary holomorphic functions of ${\zeta\equiv x+{\rm i}\, y}$. Note that for ${G\not=0}$, the  Bach tensor vanishes, while the Ricci tensor does not.

\begin{enumerate}
\item Case $p_r=0$:
All the seed metrics \eqref{pp} with \eqref{Bach0H} thus generate explicit solutions to quadratic gravity \eqref{EqQG4d} with \eqref{lambdaspec}, using the conformal transformations \eqref{solOmegaspec} and \eqref{sn_sol}  with ${p_r=0\,}$, i.e. 
$ z = p_u\, u+p_x\, x+p_y\, y=p_u\, u + p_\zeta\, \zeta+p_{{\bar\zeta}}\, {\bar\zeta}$:
\begin{enumerate}
\item
The solutions   corresponding to \eqref{solOmegaspec} are of the Weyl and traceless Ricci type~N and represent AdS waves, cf. \cite{Gullu2011}, which have the Siklos geometry \cite{GriffithsPodolsky:2009}
\BEA
\d {\tilde s}^2&=& -\frac{1}{L} \, \frac{1}{(z-c)^2}\nonumber\\
 && \times\left[ 2\left(F(\zeta)+\bar F(\bar\zeta)+ \bar\zeta\, G(\zeta)+\zeta\, \bar G(\bar\zeta)\right)\, \d u^2 -2\,\d u \d r+ \d x^2+\d y^2\right] \,. \label{siklos}
\EEA
\item
The solutions  corresponding to \eqref{sn_sol} are of the Weyl type N and traceless Ricci type~II
\BEA
\d {\tilde s}^2&=&  \sqrt{\frac{K}{L}} 
                \, {\rm sn}^2 \big( K^{\frac{1}{4}}L^{\frac{1}{4}} (z-c), -1 \big)\nonumber\\
 &&   \times      \left[2\left(F(\zeta)+\bar F(\bar\zeta)+ \bar\zeta\, G(\zeta)+\zeta\, \bar G(\bar\zeta)\right)\, \d u^2 -2\,\d u \d r+ \d x^2+\d y^2\right]\,. \label{WeylNRicciII}
\EEA
 They involve curvature singularities at zeroes of the function sn (that is at ${z=c}$ and ${z=c+P}$ where $P$ is the period) since
\BE
{\tilde R}_{ab} {\tilde R}^{ab} = \frac{1}{12} {\tilde R}^{2}\left(3 +
{{\rm sn}^{-8} \big( K^{\frac{1}{4}}L^{\frac{1}{4}} (z-c), -1 \big)}
\right).
\EE
\end{enumerate}

\item Case $p_r\not=0$:
In the exceptional case of a flat background (${2H=1}$), it is possible to employ the conformal transformation \eqref{sn_sol} with ${p_r \not=0}$. The resulting solution of quadratic gravity \eqref{EqQG4d} with \eqref{lambdaspec}
\BE
\d {\tilde s}^2=\sqrt{\frac{K}{L}} 
\, {\rm sn}^2 \big( K^{\frac{1}{4}}L^{\frac{1}{4}} (z-c), -1 \big)\left[\d u^2 -2\,\d u \d r+ \d x^2+\d y^2\right]\, \label{nonKundtmetric}
\EE
 is a conformally flat Robinson--Trautman metric. The generic case is of general Ricci type. However, in the frame 
\BE
\bl={  -}\d u\,
, \ \    \bn={  \Omega(z)^2(\d r-\pul \d u)\,, }\ \  
 \mbox{\boldmath{$m^{(2)}$}}=\Omega(z)\d x\,, \ \
\mbox{\boldmath{$m^{(3)}$}}=\Omega(z)\d y\,,   \label{frameII}
\EE
components of the Ricci tensor $R_{ab}n^a n^b$ and $R_{ab}n^a m_{(i)}^b$ vanish for 
\BE
p_r+2p_u=0\,. \label{conRTsubcase}
\EE
 Thus, in this case the vector $\bn$ is a multiply aligned null direction of the Ricci tensor and the spacetime is therefore of (traceless) Ricci type~II. 
 In fact, $\bn$ is geodetic, shearfree, twistfree, and nonexpanding 
and thus this spacetime belongs to {\it both} the Kundt (with respect to $\bn$) and Robinson--Trautman (with respect to $\bl$) classes.
\end{enumerate}

Obviously, a coordinate freedom can be used to simplify the above metrics. This is left for future work, including their physical and geometrical study.

\subsection{Robinson--Trautman solution of a general Ricci type}
As another seed metric, let us consider direct-product Kundt metrics of the form
\BE
\d s^2_{\rm seed}=2 H(r)\, \d u^2-2\,\d u\d r+\d x^2+\d y^2\,,
\EE
cf. \eqref{SSCFKundt}. For a particular function
\BE
H(r)=-\frac{1}{16}c^3 r^3 \pm  \frac{1}{4}c \sqrt{3 c d}\, r^2 - d\, r + a\,,\ \ {  cd \geq 0}\,,
\EE
the Bach tensor vanishes while the Ricci scalar is ${R=2H_{,rr}}$.
One can show that for ${a=0=d}$, the metric is conformal to a Ricci-flat spacetime  (with $\Omega=r^{-1}$), i.e. 
the metric is conformally Einstein, and from now on we study this case. 

Equation \eqref{eqOmR4d} then reads
\BE
\Omega''+\frac{3}{r}\,\Omega'+\frac{1}{r^2}\,\Omega+\frac{4{\tilde R}}{3c^3 r^3}\,\Omega^3=0\,. \label{boxORT}
\EE
An exact solution of this equation of the form 
\BE
\Omega^2(r) = \pul 
 c_1 \,r \ \ \ \mbox{with}\ \ c_1=-\frac{27c^3}{8{\tilde R}}{  >0}
\EE
leads, introducing ${\tilde r}=c_1(r/2)^2$ and rescaling both $x$ and $y$ by a constant factor ${(c_1/c)^{\frac{3}{4}}}$, to a Robinson--Trautman solution of quadratic gravity in the form
\BE
\d {\tilde s}_{\rm RT}^2=-b^2 {\tilde r}^2\,\d u^2-2\,\d u\d {\tilde r}
+\sqrt{b\,{\tilde r}}\,(\d {\tilde x}^2+\d {\tilde y}^2)\,, \label{RTgen}
\EE
where $b^2=c^3/c_1{  >0}$, {  i.e. ${\tilde R}<0$}. This metric was discussed in different coordinates in the context of conformal gravity (i.e., for ${\beta'=0=\gamma'}$) in \cite{Liu2013}, see eq.~(12) therein with $b=z_3=4 z_1= 4 z_2$). Note that this spacetime is of {\it Weyl type D} while it is of the {\it general Ricci type} (with respect to $\ell_a$d$x^a=-$d$u$).

\section*{Acknowledgments}
The authors are grateful to Marcello Ortaggio for useful comments.
 The authors acknowledge support from the
Albert Einstein Center for Gravitation and Astrophysics, Czech Science Foundation GACR
14-37086G. AP  and VP also acknowledge support from RVO: 67985840.


\end{document}